 \documentclass[final,5p,times,twocolumn]{elsarticle}

\usepackage{amssymb}
\usepackage{amsmath}
\usepackage{amsthm}
\usepackage{xcolor, color, framed}
\usepackage{ulem}
\usepackage[colorlinks, linkcolor=red,anchorcolor=green,citecolor=blue]{hyperref}

\hypersetup{
  pdftitle={Two-Pion Exchange Paradigm in Hadron-Hadron Interactions from Lattice QCD},
  pdfauthor={Tetsuo Hatsuda}
}
\usepackage{tikz,xcolor}
\usepackage{booktabs}

\newcommand{\graphsize}{0.85}

\graphicspath{{./figures/}}

\journal{Journal of Subatomic Particles and Cosmology}

\begin{document}

\begin{frontmatter}



\title{Two-Pion Exchange Paradigm in Hadron-Hadron Interactions from Lattice QCD}


\fntext[coll]{for HAL QCD Collaboration}
\author[first,second]{Tetsuo Hatsuda\fnref{coll}}
\ead{thatsuda@riken.jp}

\affiliation[first]{organization={RIKEN Center for Interdisciplinary Theoretical and Mathematical Sciences (iTHEMS)},
            addressline={Wako, 351-0198, Japan}}
\affiliation[second]{organization={Kavli Institute for the Physics and Mathematics of the Universe (Kavli IPMU), WPI, 
The University of Tokyo}, 
            addressline={Kashiwa, 277-8568, Japan}}

\begin{abstract}
We study hadron-hadron interactions in (2+1)-flavor lattice QCD near the physical pion mass ($m_\pi \simeq 146\mathrm{MeV}$) using the HAL QCD method, with particular emphasis on two-pion exchange (TPE). The nucleon--flavor-singlet hadron potentials such as $\phi$-$N$, $J/\psi$-$N$, and $\eta_c$-$N$ are found to be attractive at all distances. Moreover, in the long-range region where one-pion exchange is forbidden, we observe lattice signatures consistent with TPE dominance. These results demonstrate that the HAL QCD method can extract universal long-distance phenomena, such as TPE, directly from nonperturbative QCD dynamics.
\end{abstract}



\begin{keyword}
Lattice QCD \sep Hadron-hadron interactions \sep Two-pion exchange potential



\end{keyword}

\end{frontmatter}




\section{Introduction}
\label{introduction}

Investigating the dynamics of interacting hadrons directly from quantum chromodynamics (QCD) is one of the central goals of contemporary nuclear and particle physics. Such studies are indispensable for understanding not only the structure of ordinary nuclei and hypernuclei, but also the properties of dense baryonic matter relevant to neutron stars and other extreme astrophysical phenomena~\cite{Baym2018}. Moreover, quantitative knowledge of hadron-hadron interactions provides key theoretical input for exploring exotic hadronic states and mesic nuclei, thereby offering insight into how color confinement and spontaneous chiral symmetry breaking are realized in vacuum and in-medium~\cite{Hayano2008}.

At present, lattice QCD provides the only systematically controlled, first-principles framework for addressing the nonperturbative regime of QCD. By combining Monte Carlo sampling with large-scale numerical simulations on advanced supercomputers, lattice QCD enables direct calculations of hadronic observables from the QCD Lagrangian itself. Over the past several decades, continuous methodological refinements together with dramatic increases in computational power have led to quantitatively reliable determinations of single-hadron properties
 (see e.g.,  the physical point calculation  corresponding to a pion mass of \( m_\pi \simeq 137 \) MeV, within (2+1)-flavor QCD on a hypercubic  lattice with the lattice spacing of \( a = 0.084 \) fm and a  volume of \( V = (8.1 \, {\rm fm})^3 \) ~\cite{Fugaku-conf}.)


About hadronic interactions, two complementary methodologies have been developed within lattice QCD. One approach is based on finite-volume spectroscopy. In this so-called direct method, one computes the discrete energy eigenvalues of two-hadron systems in a finite box and relates them to infinite-volume scattering observables through L\"{u}scher’s finite-volume formalism~\cite{FVM1986}. The extraction relies primarily on the temporal behavior of two-hadron correlation functions. The second strategy is the HAL QCD method~\cite{HAL1,HAL2,HAL3,HAL4,HALReview}. Rather than focusing on discrete energy levels, this framework analyzes space-time correlations of two hadrons to define an energy-independent, generally non-local potential. The formal basis of this construction can be traced back to the Haag-Nishijima-Zimmermann reduction formula applied to four-point correlation functions of composite operators (see~\cite{Reduction} and references therein). Once the potential is determined, infinite-volume observables such as binding energies and phase shifts can be obtained by solving a Lippmann-Schwinger type integro-differential equation.

In the present article,  some hadron-hadron interactions
calculated in (2+1)-flavor lattice QCD 
with  \( a = 0.0846 \) fm, a volume of \( V = (8.12 \, {\rm fm})^3 \) 
and the pion mass  $m_\pi = 146.4$ MeV are
 presented together with its comparison to the experimental data.
 Underlying gauge configurations  were
  obtained by the 11 PFlops K computer at RIKEN \cite{K-conf}, and is called the ``K-configuration" below.

\section{The HAL QCD Method}

The HAL QCD approach formulates hadron--hadron interactions in terms of space--time correlation functions. Its basic object is the four-point correlation function for two hadrons, $h_1$ and $h_2$,
\begin{eqnarray}
F(\vec{r}, t) &\equiv & 
\langle 0 | T \Bigl[
\sum_{\vec{x}} 
\hat{h}_1(\vec{x}+\vec{r}, t)\,
\hat{h}_2(\vec{x}, t)\,
\mathcal{\hat{J}}^{\mathrm{src}}_{h_1 h_2}(0)
\Bigr] | 0 \rangle , \\
& = &  \sum_n A_n \psi_n(\vec{r}) e^{-E_n t}
+ \cdots ,
\end{eqnarray}
where $\hat{h}_1$ and $\hat{h}_2$ are interpolating (sink) operators at Euclidean time $t>0$, and 
$\hat{\mathcal{J}}^{\mathrm{src}}_{h_1 h_2}$ is a source operator inserted at $t=0$.  Also, the overlap factors 
$A_n = \langle h_1 + h_2, E_n | 
\hat{\mathcal{J}}^{\mathrm{src}}_{h_1 h_2}(0) | 0 \rangle$. 
The ellipsis denotes inelastic contributions. 
The function $\psi_n(\vec{r})$ is the equal-time Nambu--Bethe--Salpeter (NBS) wave function.

To remove trivial single-hadron propagation effects, the normalized quantity,
$\mathcal{R}(\vec{r}, t)
\equiv 
{F(\vec{r}, t)}/\left(C_{h_1}(t)\,C_{h_2}(t)\right)$,
is introduced, where $C_{h_i}(t)$ are the corresponding two-point correlation functions. 
Using $C_{h_i}(t) \sim Z_{h_i} e^{-m_{h_i} t}$ at large $t$, 
$\mathcal{R}(\vec{r}, t)$ can be written as
\begin{equation}
\mathcal{R}(\vec{r}, t)
=  \sum_n 
\frac{A_n}{Z_{h_1} Z_{h_2}}
\psi_n(\vec{r})
e^{-(E_n - m_{h_1}-m_{h_2}) t}
+ \cdots .
\end{equation}

A central result of the formalism is that the elastic part of 
$\mathcal{R}(\vec{r}, t)$ satisfies an integro-differential equation,
\begin{eqnarray}
\label{eq:NBS-eq-rewrite}
\nonumber 
\left[
\frac{1 + 3\delta^2}{8\mu}
\frac{\partial^2}{\partial t^2}
- \frac{\partial}{\partial t}
- H_0
+ O(\delta^2 \partial_t^3)
\right]
\mathcal{R}(\vec{r}, t) \\ 
= \int d^3 r'\,
U(\vec{r}, \vec{r}')
\mathcal{R}(\vec{r}', t),
\end{eqnarray}
where 
$\mu = \frac{m_{h_1} m_{h_2}}{m_{h_1}+m_{h_2}}$ 
is the reduced mass, $H_0 = -\nabla^2/(2\mu)$ is the free Hamiltonian, and 
$\delta = \frac{m_{h_1}-m_{h_2}}{m_{h_1}+m_{h_2}}$ 
quantifies the mass asymmetry. 
For the systems considered in this work, 
the higher-order time-derivative terms 
$O(\delta^2 \partial_t^3)$ are numerically negligible.

Below the inelastic threshold $E_{\mathrm{th}}$, 
the non-local kernel $U(\vec{r},\vec{r}')$ 
encodes the correct scattering phase shifts through the asymptotic behavior of the NBS wave function. 
Importantly, Eq.~(\ref{eq:NBS-eq-rewrite}) does not require isolation of the lowest eigenstate. 
All elastic states contribute consistently, so reliable results are obtained already at moderate Euclidean times,
$t \gtrsim O(\Lambda_{\mathrm{QCD}}^{-1}) \sim O(1)\ \mathrm{fm}$.

In low-energy applications, 
$U(\vec{r},\vec{r}')$ is typically approximated by a derivative expansion, $
U(\vec{r}, \vec{r}')
= \sum_n 
V_n(\vec{r}) 
\nabla^n 
\delta(\vec{r}-\vec{r}')$.
For instance, in the spin-singlet channel, 
the leading-order (LO) truncation gives a local potential,
\begin{equation}
\label{eq:V0-rewrite}
V_0^{\mathrm{LO}}(r)
=
\mathcal{R}^{-1}(r,t)
\left[
\frac{1 + 3\delta^2}{8\mu}
\frac{\partial^2}{\partial t^2}
- \frac{\partial}{\partial t}
- H_0
\right]
\mathcal{R}(r,t).
\end{equation}
The residual $t$ dependence of the right-hand side 
serves as an estimate of the systematic uncertainty 
associated with truncating the derivative expansion.
Alternatively, instead of relying on the derivative expansion, 
one may determine the full non-local potential 
$U(\vec{r},\vec{r}')$ 
by directly inverting Eq.~(\ref{eq:NBS-eq-rewrite}). 
This has recently been realized using deep neural networks, 
where the potential is optimized by minimizing the loss function
$
\mathcal{L}
=
\sum_{\vec{r},t}
\left|
(\mathrm{LHS}) - (\mathrm{RHS})
\right|^2$,
with appropriate regularization~\cite{Wang2025}. 
Here, ``LHS'' and ``RHS'' denote the left- and right-hand sides of Eq.~(\ref{eq:NBS-eq-rewrite}), respectively.

\section{$\phi$--$N$ interaction}

\begin{figure}[b]
    \centering
    \includegraphics[width=\graphsize\linewidth]{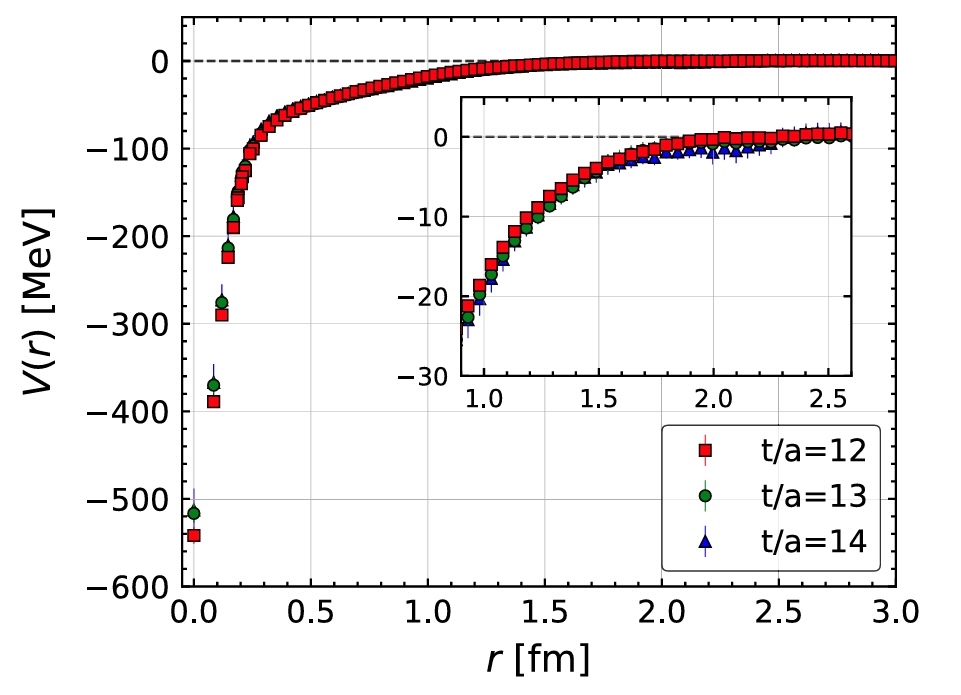}
    \caption{The $\phi$--$N$ potential $V(r)$ in the ${}^4S_{3/2}$ channel at Euclidean times
     $t/a = 12, 13, 14$~\cite{Lyu:2022imf}. }
 \label{Fig.4}
\end{figure}

Whether the $\phi$ meson can form a bound state with a nucleon has been debated for many years. 
Clarifying the nature of the $\phi$--$N$ interaction is therefore a prerequisite for addressing the possible existence of $\phi$-mesic nuclei. 
Recently, the ALICE Collaboration extracted the $\phi$--$p$ correlation function $C(k^*)$ in relative momentum space from $pp$ collisions at the LHC and reported evidence for a sizable attractive interaction~\cite{ALICE2021}. 

Motivated by this development, the HAL QCD Collaboration performed the first lattice QCD investigation of the $\phi$--$N$ system, concentrating on the highest total-spin channel, ${}^4S_{3/2}$~\cite{Lyu:2022imf}. 
Here we adopt the spectroscopic notation ${}^{2s+1}L_J$, where $s$ denotes total spin, $L$ the orbital angular momentum, and $J$ the total angular momentum. 
The ${}^4S_{3/2}$ channel is particularly advantageous because couplings to open two-body channels such as $\Lambda K ({}^2D_{3/2})$ and $\Sigma K ({}^2D_{3/2})$ are suppressed at low energies due to their $D$-wave character. 
In addition, multi-particle decay channels (e.g.\ $\Sigma \pi K$, $\Lambda \pi K$, $\Lambda \pi \pi K$) suffer from restricted phase space and are expected to contribute only weakly near threshold.

The lattice-extracted potential in the ${}^4S_{3/2}$ channel, shown in Fig.~\ref{Fig.4}, is attractive over the entire range of inter-hadron separations. 
The corresponding scattering length is qualitatively compatible with that inferred from the ALICE correlation data. 
The radial structure exhibits two characteristic features: a short-distance attractive core and a long-range attractive tail. 
A similar pattern was previously observed in the $N\Omega({}^5S_2)$ interaction~\cite{Iritani2019}. 
In both systems, quark-level Pauli blocking is absent because neither the $\phi$ nor the $\Omega$ shares valence quarks with the nucleon.

Building upon these lattice results, the experimental $\phi$--$p$ correlation function has been reanalyzed by incorporating constraints on the spin-$3/2$ interaction obtained from lattice QCD~\cite{Chizzali2024}. 
This combined analysis enabled, for the first time, an extraction of the spin-$1/2$ channel using a phenomenological complex potential guided by lattice input. 
The real part of the reconstructed potential is attractive, indicating the possibility of a $\phi$--$p$ bound state. 
The estimated binding energy in the spin-$1/2$ channel falls within the range $12.8$--$56.1$ MeV.

These results illustrate how femtoscopic correlation measurements, when supplemented by lattice QCD constraints, can provide an alternative avenue to search for hadronic bound states beyond traditional invariant-mass analyses.

\section{$J/\psi$--$N$ and $\eta_c$--$N$ interactions}

Interactions between charmonium states and the nucleon play an important role in several contexts of hadron and nuclear physics. 
The observation of the first hidden-charm pentaquark candidate $P_c$ in the $J/\psi$-$p$ channel by the LHCb Collaboration~\cite{LHCb2015} highlighted the need for reliable information on the $c\bar{c}$--$N$ force, which is an essential ingredient in coupled-channel analyses of such exotic states. 
Furthermore, the possible formation of charmonium--nucleus bound systems has been discussed extensively (see~\cite{Krein2017} and references therein), though quantitative predictions vary widely due to uncertainties in the underlying interaction.

\begin{figure}[h]
    \centering
    \includegraphics[width=\graphsize\linewidth]{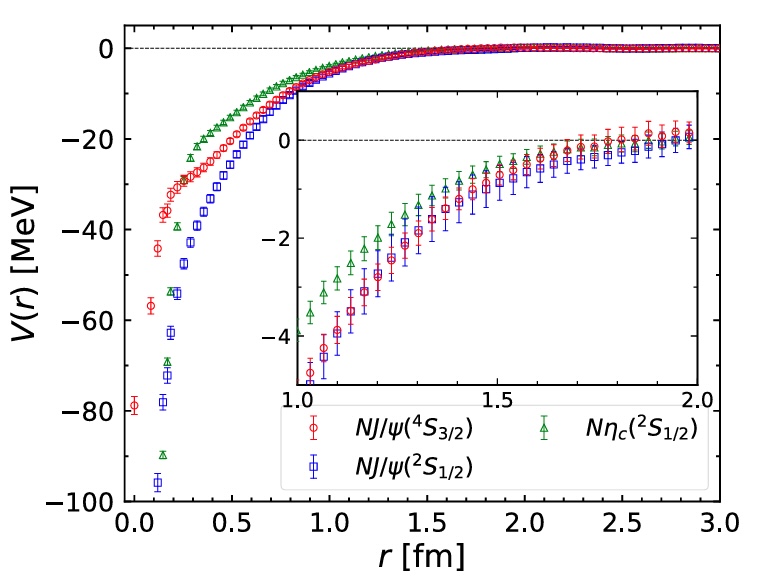}\\
    \vspace{2mm}
  \includegraphics[width=\graphsize\linewidth]{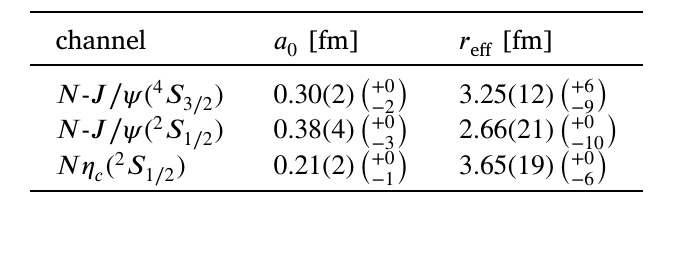}
    \caption{The $S$-wave $c\bar{c}$--$N$ potentials at $t/a = 14$ for $J/\psi$--$N$ in the ${}^4S_{3/2}$ and ${}^2S_{1/2}$ channels, and for $\eta_c$--$N$ in ${}^2S_{1/2}$. 
    The resulting scattering lengths $a_0$ and effective ranges $r_{\mathrm{eff}}$ 
     are also tabulated with statistical (first parentheses) and systematic (second parentheses) uncertainties~\cite{Lyu:2024ttm}.}
    \label{Fig.5}
\end{figure}

Beyond spectroscopy, the $c\bar{c}$--$N$ interaction contributes to the total $J/\psi$--$N$ cross section, which is relevant for understanding $J/\psi$ suppression in nuclear collisions and for constraining the intrinsic charm content of the nucleon. 
The corresponding scattering length also provides information about medium modifications of the $J/\psi$ mass in nuclear matter.

A realistic lattice QCD study of low-energy $c\bar{c}$--$N$ interactions, including both $J/\psi$ and $\eta_c$, has been carried out using the K-configuration with a physical charm quark mass~\cite{Lyu:2024ttm}. 
The resulting potentials, shown in Fig.~\ref{Fig.5}, are attractive at all distances in the $S$-wave channels considered.  From these potentials, the scattering lengths $a_0$ and effective ranges $r_{\mathrm{eff}}$ were determined as shown in Fig.~\ref{Fig.5}. 
Future femtoscopic analyses of $J/\psi$--$p$ correlations at the LHC will provide an opportunity to test these quantitative predictions.

Using the spin-averaged scattering length 
$
a_{J/\psi}^{\mathrm{spin\text{-}av}}
=
(2a_0^{(3/2)} + a_0^{(1/2)})/3$,
one can estimate the in-medium mass shift of the $J/\psi$ under the linear-density approximation,
\begin{equation}
\delta m_{J/\psi}
\simeq
-\frac{2\pi (m_N + m_{J/\psi})}
     {m_N m_{J/\psi}}
a_{J/\psi}^{\mathrm{spin\text{-}av}}
\rho_0
=
-19(3)\ \mathrm{MeV},
\end{equation}
where $\rho_0 = 0.17~\mathrm{fm}^{-3}$ is the normal nuclear matter density. 
Such an attractive mass shift suggests that charmonium--nucleus bound states may exist for sufficiently heavy nuclei, providing promising targets for future experimental searches~\cite{Krein2017}.

\section{Two-pion Exchange (TPE) Paradigm}

\subsection{TPE in the Nuclear Force}
\label{sec:TPE-NN}

The long- and intermediate-range structure of the nucleon--nucleon ($NN$) interaction is governed by pion dynamics. 
At distances of order $r \sim m_\pi^{-1}$, the dominant contribution arises from one-pion exchange (OPE), as originally proposed by Yukawa~\cite{Yukawa:1935xg}. 
When the separation decreases to $r \sim (2m_\pi)^{-1}$, mechanisms involving the exchange of two pions become increasingly important. 
This intermediate-range attraction was already emphasized in the pioneering work of Taketani, Machida, and Ohnuma (the TMO potential)~\cite{TMO:1951}.

\begin{figure}[t]
    \centering
    \includegraphics[width=9cm]{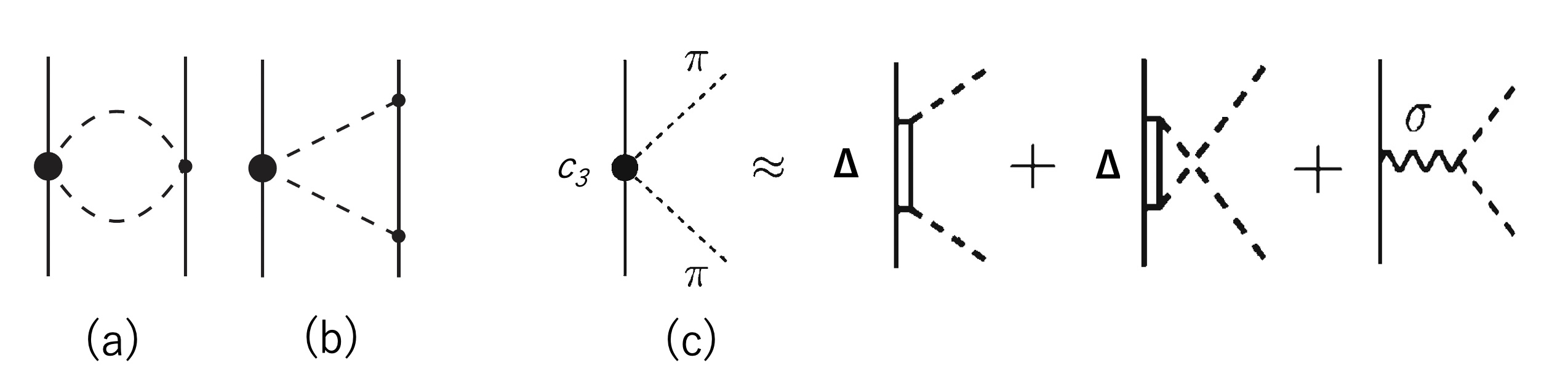}
    \caption{(a,b) Football and triangle diagrams contributing to the $NN$ TPE potential at N$^2$LO. 
       (c) Meson-theoretic interpretation of the $c_3$ vertex in $\widehat{\mathcal{L}}^{(2)}_{\pi N}$, 
    involving $\Delta$ excitation and $S$-wave $\pi\pi$ correlation ($\sigma$).}
    \label{Fig6}
\end{figure}

Within modern chiral effective field theory (EFT), two-pion exchange (TPE) contributions are organized systematically in powers of momenta and the pion mass. 
While TPE diagrams start to appear at next-to-leading order (NLO), the spin- and isospin-independent central component $V_{\rm C}$ emerges only at next-to-next-to-leading order (N$^2$LO); see, e.g., Refs.~\cite{Machleidt:2011zz,Epelbaum:2008ga}. 
At this order, the relevant topologies are the football and triangle diagrams shown in Fig.~\ref{Fig6}(a,b),
 where the large (small) solid dot represents a vertex from the second-order (first-order) $\pi N$ Lagrangian, 
    $\widehat{\mathcal{L}}^{(2)}_{\pi N}$ ($\widehat{\mathcal{L}}^{(1)}_{\pi N}$) \cite{Machleidt:2011zz}.
  (See the Appendix A for the explicit forms). 
Interestingly, the football diagram does not generate a central contribution, so the triangle diagram alone determines the state-independent TPE potential.

In the heavy-nucleon limit, the dominant TPE contribution to the central force arises from the interference of the $c_3$ vertex in $\widehat{\mathcal{L}}^{(2)}_{\pi N}$ with the axial coupling $g_A$ from $\widehat{\mathcal{L}}^{(1)}_{\pi N}$.
Microscopically, the constant $c_3$ reflects $\pi N$ dynamics involving intermediate $\Delta$ excitation and correlated $S$-wave $\pi\pi$ exchange (often associated with the $\sigma$ channel)~\cite{Kaiser:1998wa,Machleidt:2020vzm}; see Fig.~\ref{Fig6}(c). 
The explicit coordinate-space expression for the state-independent TPE central potential at N$^2$LO reads,
$\left. V_{\rm NN;C}^{2\pi}(r) \right|_{\rm N^2LO}^{\rm Fig.\,\ref{Fig6}(a)}  = 0$ and 
$ \left. V_{\rm NN;C}^{2\pi}(r) \right|_{\rm N^2LO}^{\rm Fig.\,\ref{Fig6}(b)} \neq 0$
with an explicit form ~\cite{Kaiser:1997mw};
\begin{equation}
 \left. V_{\rm NN;C}^{2\pi}(r) \right|_{\rm N^2LO}
=
\frac{3g_A^2}{32\pi^2 f_\pi^4}
\frac{e^{-2x}}{r^6}
\left\{ 2 c_1f_1(x) + c_3 f_3(x)
\right\}.
\end{equation}
where $x = m_\pi r$,
 $f_1(x)=  x^2 (1+x)^2$ and  $f_3(x)= 6 + 12x + 10x^2 + 4x^3 + x^4$.
 Empirical values of the low-energy constants are $c_1 = -1.10(3)$ and  $c_3 = -5.54(6)$,
 in the unit of GeV$^{-1}$ (Appendix A). 

Two limiting behaviors are particularly illuminating. 
For $x \ll 1$ (formally corresponding to a vanishing pion mass), one finds
\begin{equation}
V_{\rm NN;C}^{2\pi}(r)
\xrightarrow[x \ll 1]{} 
\frac{9 g_A^2 c_3}{16\pi^2 f_\pi^4}
\frac{1}{r^6} ,
\label{vanderW-rewrite}
\end{equation}
which has the structure of a van der Waals interaction. 
In contrast, at large distances with a fixed pion mass, $x \gg 1$,
\begin{equation}
V_{\rm NN;C}^{2\pi}(r)
\xrightarrow[x \gg 1]{} 
\frac{3 g_A^2 c_3 m_\pi^4}{32\pi^2 f_\pi^4}
\frac{e^{-2m_\pi r}}{r^2} ,
\label{yukawa2-rewrite}
\end{equation}
demonstrating that asymptotically the TPE contribution behaves like the square of the one-pion exchange potential.

\subsection{TPE between nucleon and flavor-singlet hadron}
\label{sec:TPE-FSN}

\begin{figure}[t]
    \centering
    \includegraphics[width=7cm]{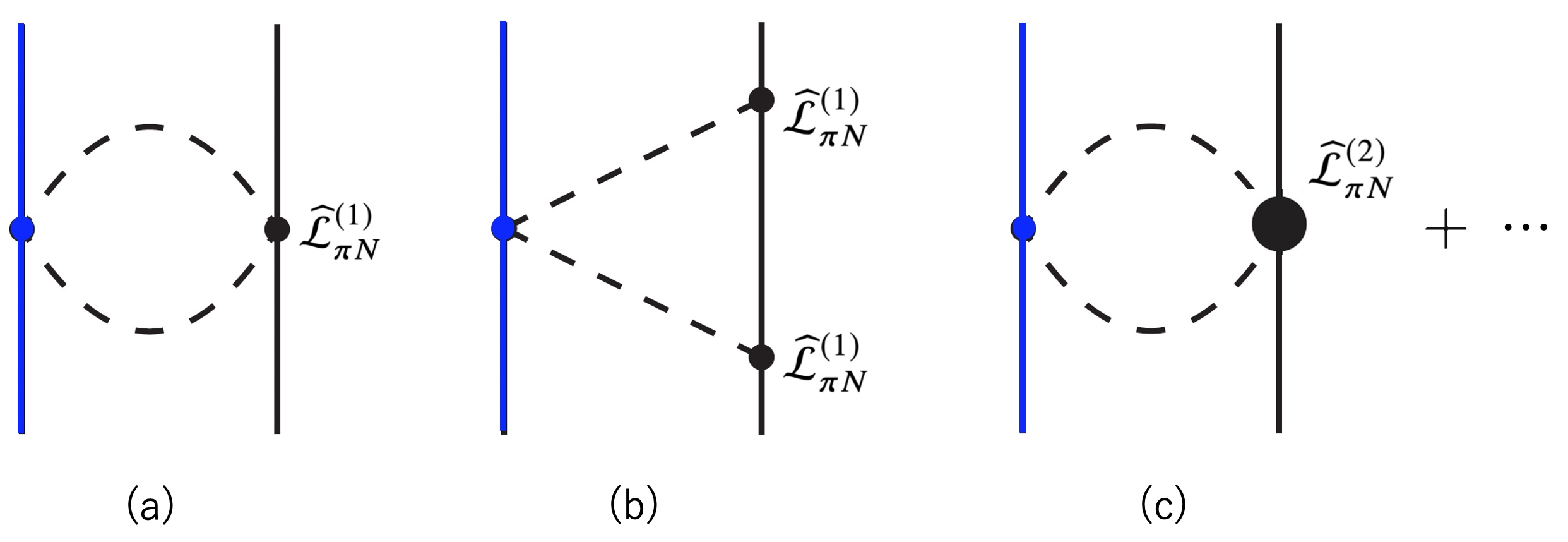}
  \caption{(a) The N$^3$LO football TPE diagram between a flavor-singlet hadron (the blue line) and the nucleon through the two-pion exchange. The blue solid dot is given by 
   Fig.\ref{Fig.7}(b), while the small 
   black solid dot is from the $\pi \pi NN$ seagull term 
   in $\widehat{\mathcal{L}}^{(1)}_{\pi N} $.  This contribution turns out to be zero due to isospin structure of the vertices \cite{TarrusCastella:2018php}.  
   (b) The N$^3$LO triangle TPE diagram with the small   black solid dot from the $\pi NN$ term
    proportional to $g_A$ in $\widehat{\mathcal{L}}^{(1)}_{\pi N} $.
    (c) The N$^4$LO contributions including the football TPE diagram with the large black solid dot from the terms
    in $\widehat{\mathcal{L}}^{(2)}_{\pi N} $.}
    \label{Fig.8}
\end{figure}

If one of the nucleons  in Sec.~\ref{sec:TPE-NN}  is replaced by a flavor-singlet hadron,
 OPE  is inhibited and the possible long-range interaction should come from TPE 
first appearing in the N$^3$LO (Fig.\ref{Fig.8}(a,b)).  These diagrams  in the chiral EFT approach
 have been  evaluated  as
 $ \left. V^{2\pi}_{Q\bar{Q}-N;{\rm C}} (r) \right|_{\rm N^3LO}^{{\rm Fig.\,\ref{Fig.8}(a)}} =0 $ and
 $ \left. V^{2\pi}_{Q\bar{Q}-N;{\rm C}} (r) \right|_{\rm N^3LO}^{{\rm Fig.\,\ref{Fig.8}(b)}}  \neq 0$ 
  \cite{TarrusCastella:2018php} with  
\begin{eqnarray}
 \left. V^{2\pi}_{Q\bar{Q}-N;{\rm C}} (r) \right|_{\rm N^3LO} 
=   \frac{3g_A^2}{128\pi^2 f_{\pi}^2} \frac{e^{-2x}}{r^6}  \Big\{ c_{{di}} g_d(x)+ c_m g_m(x) \Big\},
   \label{full-2}
  \end{eqnarray} 
 where $x=m_{\pi}r$, $g_d(x)= 6 + x (2 + x)(6 + x (2 + x))$, $g_m(x) = x^2 (1 + x)^2$,
$c_{di} = -4\pi^2\beta\kappa_2/b$, and $c_m = -12\pi^2\beta/b$.
Here $b = 11 - 2N_f/3$, $\beta$ is the color-electric polarizability of the quarkonium ground state,
and $\kappa_{1,2}$ describe the two-pion coupling to the chromoelectric field.
Its asymptotic form for $x \gg 1$ reads
\begin{eqnarray}
 \left. V^{2\pi}_{Q\bar{Q}-N;{\rm C}} (r) \right|_{\rm N^3LO} 
 \xrightarrow[x \gg 1]{} 
 \frac{3g_A^2 (c_{{di}} + c_m) m_\pi^4 }{128\pi^2 f_{\pi}^2 } \frac{e^{-2m_\pi r}}{r^2}.
 \label{asymp-2}
\end{eqnarray}
 Note, however,  that the asymptotic form Eq.(\ref{asymp-2}) significantly underestimates 
 Eq.(\ref{full-2}) at distances $r \sim (2m_{\pi})^{-1}$.

In Fig.~\ref{Fig.8}(c), the contributions at N$^4$LO are shown, highlighting the 
football diagram that includes the $\pi\pi NN$ vertex from $\widehat{\mathcal{L}}^{(2)}_{\pi N}$.
Although this diagram is formally of higher order compared to the triangle diagram 
in Fig.~\ref{Fig.8}(b), it may lead to an enhancement, similar to the case of TPE in the $NN$ interaction 
$\left. {V}_{\rm NN;C}^{2\pi}(r) \right|_{\rm N^2LO}^{\rm Fig.\,\ref{Fig6}(b)}$ occurring through the $c_3$-term, driven by the $\Delta$ excitation. Further details of this analysis will be reported elsewhere \cite{HHLWW2016}.

\subsection{Signatures of TPE in $\phi$-N, $J/\psi$-N and $\eta_c$-N systems}
\label{sec:TPE-lattice}

To examine whether the TPE mechanism between a flavor-singlet hadron and a nucleon, discussed in Sec.~\ref{sec:TPE-FSN}, is realized in practice, we analyze the lattice-QCD potentials $V_{\rm lat}(r)$ for the $\phi$--$N$, $J/\psi$--$N$, and $\eta_c$--$N$ systems reported in Refs.~\cite{Lyu:2022imf,Lyu:2024ttm}.  For this purpose, we introduce a ``spatial effective energy'' defined as a function of the separation $r$,
\begin{eqnarray}
E_{\text{eff}}(r)
=
-\frac{\ln\!\left[-V_{\rm lat}(r)\, r^{\nu}/\alpha\right]}{r}.
\end{eqnarray}
If TPE indeed governs the interaction at distances $r>(2m_\pi)^{-1}$, then $E_{\text{eff}}(r)$ should approach the expected value $2m_\pi=292.8~\text{MeV}$ for the K-configuration.

\begin{figure}[t]
    \centering
    \includegraphics[width=\graphsize\linewidth]{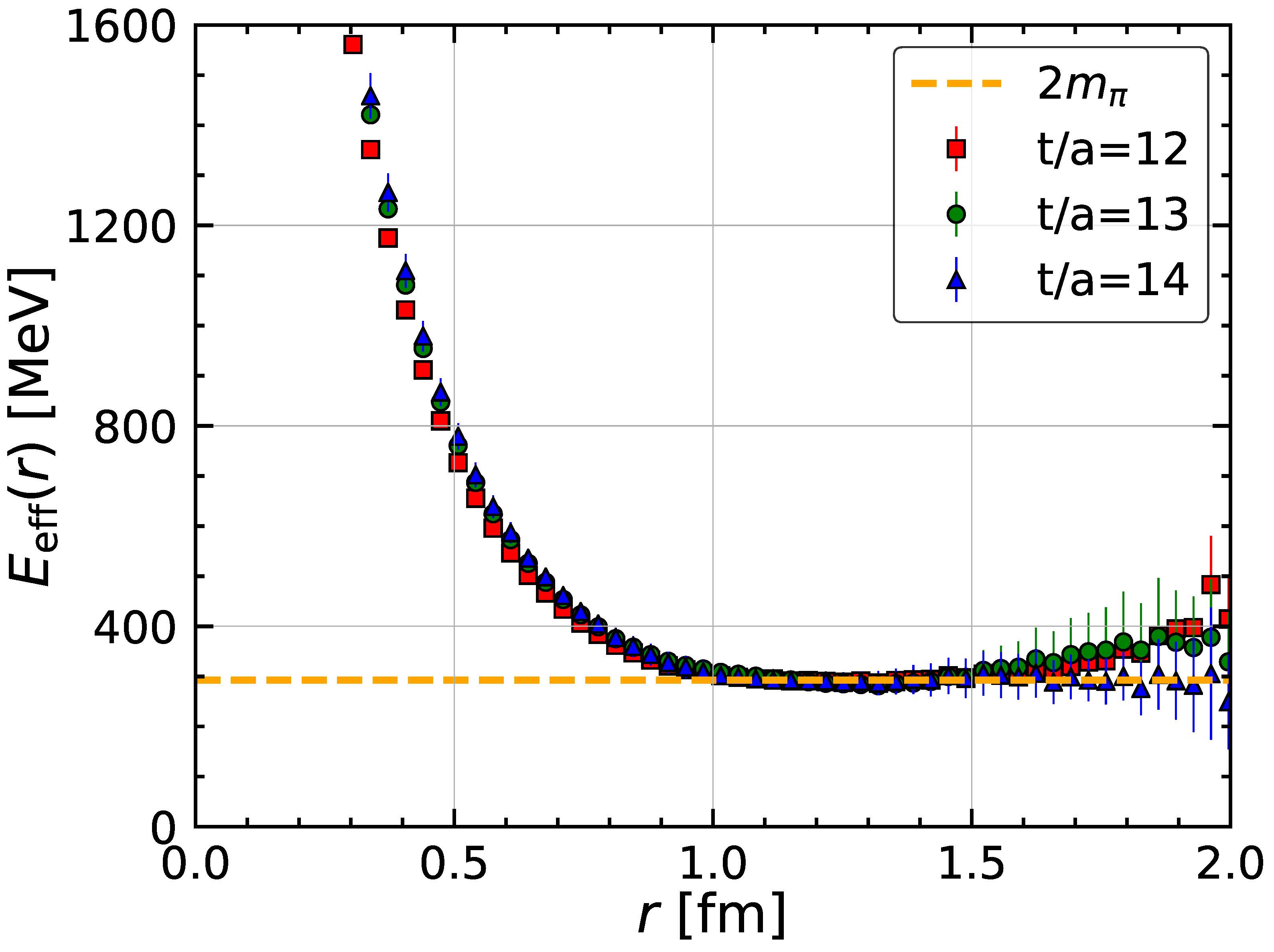}\\
    \includegraphics[width=\graphsize\linewidth]{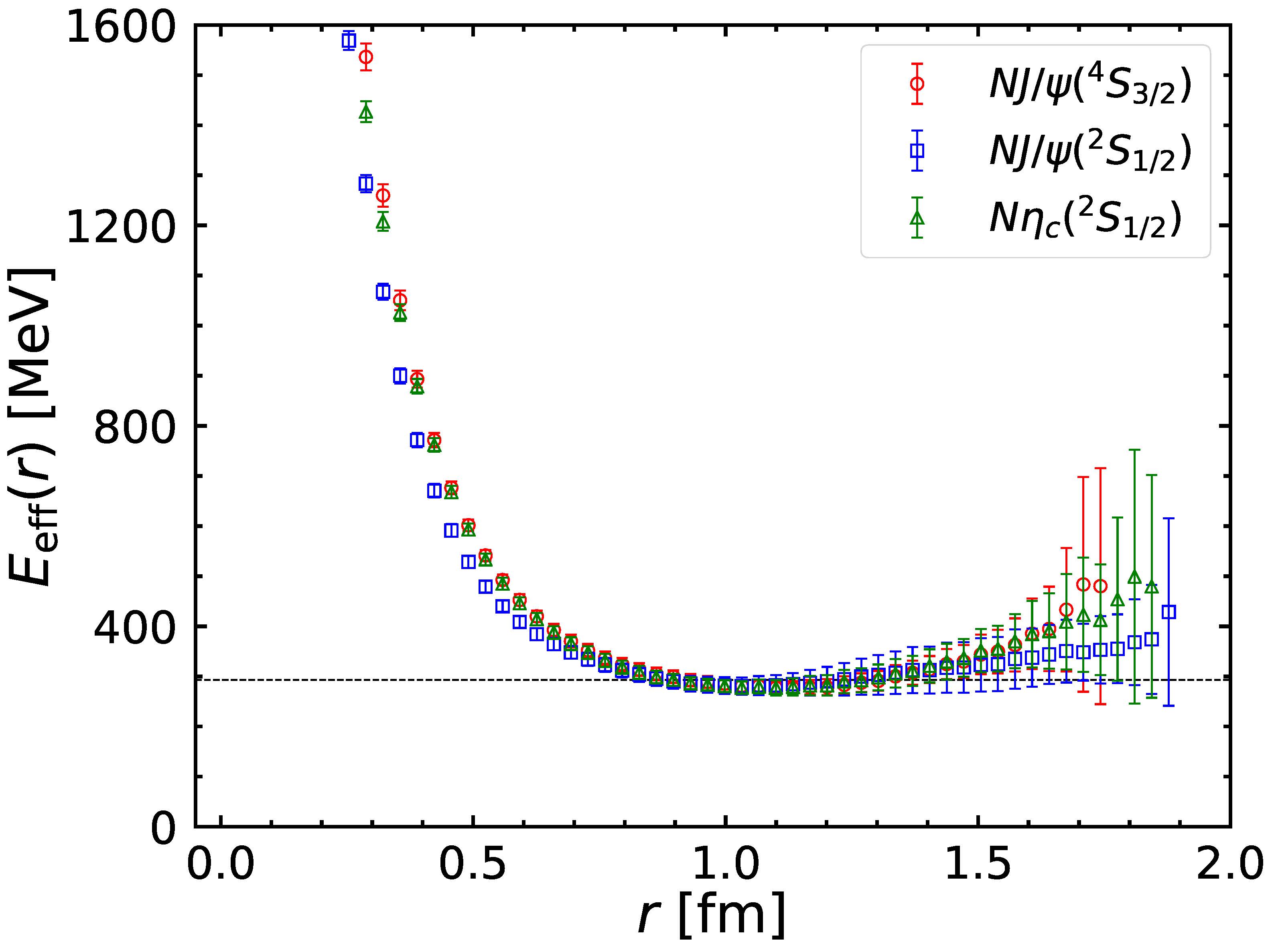} 
   \caption{(Upper panel) 
   The spatial effective energy $ E_{\text{eff}}(r) $ as a function of separation $ r $ at Euclidean time 
   $ t/a = 12 $ (red squares), 13 (green circles), and 14 (blue triangles). The orange dashed line corresponds to $ 2m_\pi $ with the known value of the pion mass $ m_\pi = 146.4 $ MeV in the K-configuration \cite{Lyu:2022imf}.
    (Lower panel)  $ E_{\text{eff}}(r) $ for 
     two spin states ($ J$ = 1/2  and 3/2 ) of the $ J/\psi -N $ system, and the
      $ J = 1/2 $ state of the $\eta_c$--$N$ system at $ t/a = 14$. The black dotted line corresponds to
      $ 2m_\pi = 292.8 $ MeV
     (unpublished results based on~\cite{Lyu:2024ttm}).
   }
    \label{Fig.9}
\end{figure}

In the present study we take $\nu=2$, guided by Eq.~(\ref{asymp-2}). 
The normalization constant $\alpha$ is fixed by fitting the lattice data at $t/a=14$ in the interval $0.9~\mathrm{fm}<r<1.8~\mathrm{fm}$. 
The upper panel of Fig.~\ref{Fig.9} shows $E_{\text{eff}}(r)$ for the $\phi$--$N$ interaction at $t/a=12,13,14$~\cite{Lyu:2022imf}. 
The lower panel displays the results for the $J/\psi$--$N$ system in the two spin channels ($J=1/2$ and $3/2$), together with the $J=1/2$ channel of the $\eta_c$--$N$ system at $t/a=14$ (unpublished results based on Ref.~\cite{Lyu:2024ttm}).

In all cases, we find that $E_{\text{eff}}(r)$ is consistent with a plateau near $2m_\pi=292.8~\text{MeV}$ for $r \gtrsim 1.0~\mathrm{fm}$ within statistical uncertainties. 
This behavior supports the interpretation that the long-range part of these interactions is dominated by the TPE potential. 
To solidify this conclusion, further progress is required, including improved lattice statistics to reduce uncertainties and a more complete theoretical treatment of the football diagram in Fig.~\ref{Fig.8}(c) within chiral EFT, as suggested in Sec.~\ref{sec:TPE-FSN}.

\subsection{TPE between flavor-singlet hadrons}
\label{sec:TPE-FSFS}

 \begin{figure}[t]
\hspace{1cm}
    \centering
    \includegraphics[width=6.5cm]{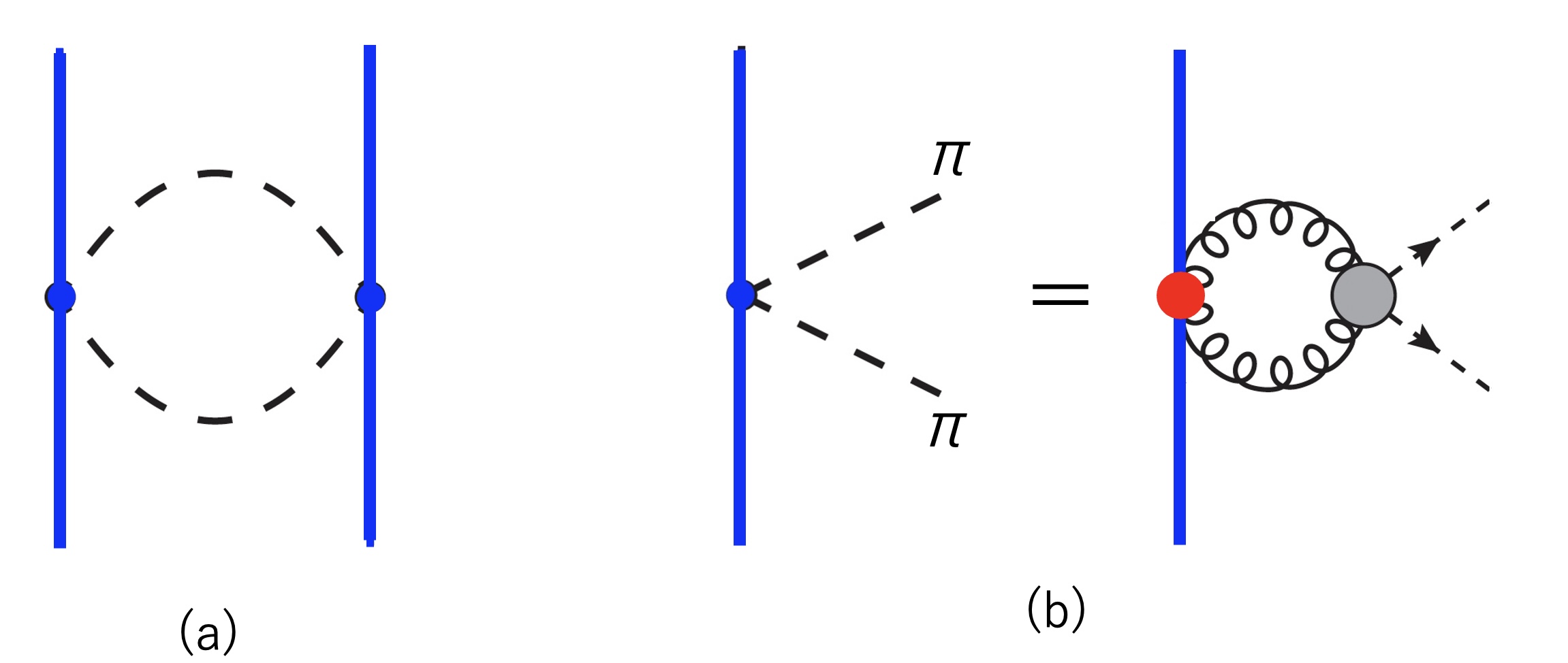}
   \caption{(a) The long-range interaction between the two flavor-singlet hadrons (the blue lines) through the two-pion exchange.  (b) The microscopic origin of the coupling of a
   flavor singlet hadron and the two pions (the solid blue dot) through
    the gluon cloud \cite{Brambilla:2015rqa}.}
    \label{Fig.7}
\end{figure} 


If, in the discussion of Sec.~\ref{sec:TPE-NN}, both nucleons are replaced by flavor-singlet hadrons, OPE is forbidden by quantum numbers, and the longest-range force necessarily originates from TPE. 
A representative case is the interaction between two heavy quarkonia, such as bottomonium--bottomonium or charmonium--charmonium systems. 
In this situation, the relevant mechanism corresponds to the football diagram shown in Fig.~\ref{Fig.7}(a). 
The gluonic fields surrounding each $(Q\bar{Q})$ bound state couple to flavor-singlet two-pion states [Fig.~\ref{Fig.7}(b)], which can then be exchanged between the color dipoles, generating the TPE potential.

This class of interactions was first analyzed by Bhanot and Peskin~\cite{Bhanot:1979vb}, and subsequently reformulated using dispersion-theoretical methods~\cite{Fujii:1999xn} and within the framework of chiral EFT~\cite{Brambilla:2015rqa}. 
An explicit expression for the central TPE potential between two heavy quarkonia was obtained in Ref.~\cite{Brambilla:2015rqa}:

 \begin{eqnarray}
 \left. V^{2\pi}_{Q\bar{Q}-Q\bar{Q};{\rm C}}(r) \right|_{\rm Fig.\,\ref{Fig.7}(a)}
 =   -\frac{3\pi \beta^2 m_{\pi}^2}{8b^2 r^5} 
 \sum_{n=1,2} g_n(x) K_n (2x) , 
 \label{full-Q}\\
 \hspace{2cm} \xrightarrow[x \gg 1]{}  -\frac{3(3+\kappa_2)^2 \pi^{3/2} \beta^2 m_{\pi}^{9/2} }{4b^2}  \cdot \frac{e^{-2m_{\pi}r}}{r^{5/2}}.
\label{asymp}
\end{eqnarray}
 Here $x=m_{\pi}r$,
 $g_1(x)=  4(\kappa_2 + 3)^2 x^3
+ (3\kappa_1^2 + 43\kappa_2^2 + 14\kappa_1 \kappa_2) x $, $g_2(x) = 2(2(\kappa_2 + 3)(\kappa_1 + 5\kappa_2)x^2 
+ 3\kappa_1^2 + 43\kappa_2^2 + 14\kappa_1 \kappa_2)$ and
 $K_n(y)$ denotes the modified Bessel function of the second kind. 
Note that the asymptotic form 
Eq.(\ref{asymp}) significantly underestimates the full result Eq.(\ref{full-Q}) at distances \( r \sim (2m_{\pi})^{-1} \).  

The potential in Eq.~(\ref{full-Q}) offers a useful starting point for analyzing lattice-QCD results on $J/\psi$--$J/\psi$ and $\eta_c$--$\eta_c$ interactions within the HAL QCD framework. 
Moreover, a generalization of this formalism to heavy baryons would be of considerable interest for studying the long-range behavior of $\Omega_{sss}$--$\Omega_{sss}$ and $\Omega_{ccc}$--$\Omega_{ccc}$ systems, for which lattice-QCD data have already been reported by the HAL QCD Collaboration~\cite{Gongyo:2017fjb,Lyu:2021qsh}.


\section{Summary}
\label{summary}

In this article, we highlight recent developments in lattice-QCD studies of hadron--hadron interactions near the physical pion mass ($m_{\pi}\simeq 146~\mathrm{MeV}$), based on the HAL QCD method. We place particular emphasis on meson--baryon systems: the extracted $\phi$--$N$, $J/\psi$--$N$, and $\eta_c$--$N$ potentials are attractive over the full range of inter-hadron separations. These results provide a foundation for investigating interactions between nucleons and flavor-singlet hadrons.

We also examine the two-pion-exchange (TPE) mechanism, which governs the long-distance behavior of interactions among flavor-singlet hadrons and between nucleons and flavor-singlet hadrons. For the latter, lattice-QCD potentials exhibit features consistent with TPE. In addition, we point out the potential importance of the $\pi\pi NN$ seagull term in the second-order chiral Lagrangian, which is known to play a leading role in TPE contributions to the $NN$ potential.

Overall, the results highlighted here demonstrate that the HAL QCD method provides a robust and quantitative framework for probing nonperturbative QCD dynamics, effectively bridging lattice predictions and experimental observations.

\section*{Acknowledgements}

The author appreciates the HAL QCD Collaboration for many valuable discussions.  Special thanks are due to Yan Lyu, Takumi Doi and Liang Zhang for their helpful discussions regarding two-pion-exchange dynamics in hadronic interactions. 
This work was supported in part by the Japan Science and Technology Agency (JST) through the ASPIRE program (Adopting Sustainable Partnerships for Innovative Research Ecosystem), Grant No.~JPMJAP2318.

\appendix

\section{Pion-nucleon vertices}
\label{App:A}

The pion-nucleon interactions involving {\it up to two pions}
 in the first- and second-order chiral Lagrangians
 of the heavy-baryon formalism read \cite{Machleidt:2011zz}
\begin{eqnarray}
& & \hspace{-0.5cm} \widehat{\mathcal{L}}^{(1)}_{\pi N} \rightarrow  \bar{N} \left\{ 
- \frac{1}{4 f_\pi^2} \boldsymbol{\tau} \cdot \left( \boldsymbol{\pi} \times \partial_0 \boldsymbol{\pi} \right) 
- \frac{g_A}{2 f_\pi} \boldsymbol{\tau} \cdot \left( \vec{\sigma} \cdot \vec{\nabla} \right) \boldsymbol{\pi} 
\right\} N, \nonumber \\
& & \hspace{-0.5cm}  \widehat{\mathcal{L}}^{(2)}_{\pi N} \rightarrow 
\bar{N} \Bigg[ 
- \frac{2c_1}{f_\pi^2} m_\pi^2 \, \boldsymbol{\pi}^2 \nonumber \\
& &  \hspace{1.0cm} + \left( c_2 - \frac{g_A^2}{8M} \right) \frac{1}{f_\pi^2} (\partial_0 \boldsymbol{\pi} \cdot \partial_0 \boldsymbol{\pi})  + \frac{c_3}{f_\pi^2} (\partial_\mu \boldsymbol{\pi} \cdot \partial^\mu \boldsymbol{\pi}) \nonumber\\
& & \hspace{1.0cm} - \left( c_4 + \frac{1}{4M} \right) \frac{1}{2f_\pi^2} 
\, \epsilon^{ijk} \epsilon^{abc} \sigma^i \tau^a 
(\partial^j \pi^b)(\partial^k \pi^c) 
\Bigg] N .
\nonumber
\end{eqnarray}
Here $M$ is the nucleon mass and 
the  empirical values of $c_{1-4}$ from 
Table 1 of \cite{Machleidt:2020vzm} read,  
 $c_{1} = -1.10(3)$,
  $c_{2} = 3.57(4)$,
 $c_{3} = -5.54(6)$,
 and  $c_{4} = 4.17(4)$,
  in the unit of ${\rm GeV}^{-1} $.



\end{document}